\documentclass[12pt]{article}
\usepackage{amsfonts}
\begin{document}
\title{Inverse eigenvalue problem
for discrete three-diagonal Sturm-Liouville operator and the
continuum limit}

\author{V. M. Chabanov and B. N. Zakhariev \\
Laboratory of theoretical physics, JINR, \\
Dubna, 141980, Russia \\
email: chabanov@thsun1.jinr.ru \\
 } \date{}

\maketitle

\begin{abstract}
In present article  the self-contained derivation of eigenvalue
inverse problem results is given by using a discrete approximation
of the Schr\"odinger operator on a bounded interval as a finite
three-diagonal symmetric Jacobi matrix. This derivation is more
correct in comparison with previous works which used only
single-diagonal matrix. It is demonstrated that inverse problem
procedure is nothing else than well known Gram-Schmidt
orthonormalization in Euclidean space for special vectors numbered
by the space coordinate index.  All the results of usual inverse
problem with continuous coordinate are reobtained by employing a
limiting procedure, including the Goursat problem -- equation in
partial derivatives for the solutions of the inversion integral
equation.
\end{abstract}
PACS
%03.65.-w,
02.30.Zz

\section{Introduction}

There is a vast literature on the inverse scattering problem.
Suffice it to mention the classical monographs
[1-5],
%\cite{AgranMar,Levit,Mar,ChSab,New},
see also \cite{FY}. That theory
has multiple applications and never stopped developing \cite{Sab}. In
parallel with its renovation, attempts to give a clear and obvious
treatment were undertaken [8-10].
%\cite{CK,C,Cn}.
In these last papers that was done
by using finite-difference approach which reduces the problem to
solving relatively simple algebraic equations.  Passage to the limit of the
continuous variable allows one to obtain classical results of the inverse
problem.  Thus, the finite-difference version represents a valuable
tool to reproduce all the results of the inversion procedure on a more
accessible level of understanding. And this is not only of pedagogical
interest. The aim of science is, among others, to supply a maximally
compacted and clear knowledge free from superfluous and
often obscure details.

The authors of [8-10]
%\cite{CK,C,Cn}
restricted their consideration to the finite-difference matrix
Hamiltonian with potential coefficients only on the main diagonal of the
operator. So, there appears a
disparity in numbers of interaction parameters and spectral data (see
discussion below). As a result, we need either to impose some restrictions
on spectral parameters or to introduce additional non-local potentials as
was done in present article. That problem was not considered in [8-10],
%\cite{CK,C,Cn},
which led to an "erroneous" final result for the potential in the
continuum limit:  there must be additional factor 2 (missed in the
articles mentioned) in front of derivative of the solution of
inverse problem integral equation [see Eq. (\ref{vK})]. This
oversight was partially compensated in the paper \cite{Berry}
where the authors introduced non-local potentials which are needed
for correct final results in the continuum limit. However, their
method had several deficits, among them one can mention a rather
"adjustable" character of the procedure suggested and none of
their algorithms indicates the uniqueness.

At the same time, the inverse eigenvalue problem for the discrete analog
of the Sturm-Liouville operator is now well developed. There is
sufficiently large number of papers on that subject, see \cite{Ges} where
one can find most of the references. There are different
variants of the inverse problem in the discrete approximation. All
they deal with a matrix (finite or infinite) with several diagonals which
may differ in number and which are recovered by given (spectral)
parameters.  For the Gel'fand-Levitan analog, the most known inversion
variant, the central theorem is valid: Given the set of eigenvalues of a
three-diagonal Jacobi matrix and the first components of the associated
orthonormal eigenvectors, there exists a unique Jacobi matrix corresponding
to these data, see e.g. \cite{Ges,GladWil} and references therein.
The usual proof of this theorem is performed using orthogonal polynomials
\cite{Ges}.

Thus, we have the continuous and discrete variants where the
 inversion procedure is well established. However, considerably less
was done to link them both. The problem is likely that the
recovery procedures in discrete and continuum cases outwardly have
little in common.  In the continuum version, the inversion
procedure (by Gel'fand-Levitan-Marchenko) is built as a transition
from a certain known system (free motion, as a rule) to the system
with known spectral data but with unknown potential to be
reconstructed. The aim of the paper is to give such a derivation
of the inverse eigenvalue problem from its discrete variant which
would be free from the previous errors.

In the discrete variant, we have to develop, in a  more explicit
form, a structure similar to that in the continuum case.  In doing
so, we have to use a general criterion which would strongly
specify whether our development is correct. The orthogonal
polynomials' method gives such a hint. This is orthonormality
relation (in a special spectral measure) which is valid for any
system.  As is well known, the Gram-Schmidt method is essential in
constructing these polynomials. So the central idea of the present
paper is to employ that method in order to reconstruct the set of
eigenvectors orthonormalized in the initial spectral measure when
the last is changed in a given way. We shall construct some
"prototype" of the transformation procedure that realizes the
reconstruction of the potential of regular Schr\"odinger operator
in the continuum case.

We begin, in the second section, with several results consisting
of some preliminary constructions which appear as intermediate
steps in the course of the methods discussed. First of all, we
give the discrete statement of the Sturm-Liouville problem on a
bounded interval with zero boundary conditions, which is
equivalent to consideration of a three-diagonal symmetric Jacobi
matrix. In the continuum limit these diagonals merge in a single
diagonal (local potential).  Then we pose the inverse eigenvalue
problem in terms of eigenvalues and associated spectral weight
factors, first components of the orthonormalized eigenvectors.
Introducing so-called regular solutions admits of the explicit
presentation of these parameters which serve as a spectral measure
(of bounded support) entering in the Parseval relation for
eigenvectors.  That measure allows us to represent this equality
as orthonormality condition for the same vectors but from another
standpoint when the energies represent components and the discrete
coordinate numbers the vectors. Next step is applying Gram-Schmidt
technique to obtain the new orthonormal set of eigenvectors (in
the sense mentioned) corresponding to the new measure. We shall
see  that procedure indeed reproduces prototypes for  equations of
the inverse problem in the continuum limit.  The proof is given
that such a orthogonalization is the only possible development.
Next, we use the new eigenvectors to recover the potential
coefficients on the three diagonals of the Jacobi matrix (discrete
Sturm-Liouville operator) by using completeness relation for the
new eigenvectors.

In the third section, we pass to the continuum limit. We
demonstrate how all the  discrete equations-prototypes go over
into the classical equations of the inverse Sturm-Liouville
problem: Gel'fand-Levitan equations, expression for the potential,
classical Goursat problem, etc. That accomplishes our program.

\section{Discrete version of inverse problem on finite interval}

It is most easy to demonstrate the essence of the inverse eigenvalue
problem for the example of finite-difference Schr\"odinger equation in the
discrete variable $x_{n}, n \in {\mathbb Z}$ with the step $\Delta$:
\begin{eqnarray}
-\frac{\Psi(x_{n+1},E)-2 \Psi(x_{n},E)+\Psi(x_{n-1},E)}{\Delta ^2}+
V(x_{n}) \Psi(x_{n},E) + \nonumber \\ u(x_{n}) \Psi(x_{n+1},E) +
u(x_{n-1}) \Psi(x_{n-1},E) = E \Psi(x_{n},E),
\label{fdsch}
\end{eqnarray}
$V(x_{n})$ and $u(x_{n})$  are real, for this problem is reduced
to linear algebraic equations.  The first three terms in this
equation represent finite-difference operator of the second
derivative, i.e. kinetic energy. Note the existence, in the
Schr\"odinger equation, of terms $u(x_{n})$ corresponding to a
"minimally non-local" interaction.  We shall soon come back to
them and their introduction will turn out justified.

Let us consider the bounded interval $[0, \pi]$ with finite number
N of points inside: $x_{0}=0; \enskip x_{N+1}=\pi$, so that
$\Delta = x_{n+1}-x_{n}=\pi/(N+1)$.  Let us add the Eq. (1) by the
Dirichlet boundary conditions:  \begin{equation}
\Psi(x_{0},E)=\Psi(x_{N+1},E)=0. \label{fdbc}
\end{equation}
These zero boundary conditions have the well-known physical
interpretation that the movement of a particle is restricted by
the infinitely tall walls at the points $x_{0}$ and $x_{N+1}$ (the
infinite rectangular potential well).

The spectrum of the problem (\ref{fdsch}), (\ref{fdbc}) is a
ladder of discrete energy levels $\{E_{\nu
}\}_{\nu=1}^{N}$ for bound states representing the unit vectors
$\Psi_{\nu }(x_{n})$ $\equiv$ $\Psi(x_{n},E_{\nu})$, $$\sum_{n=0}^{N+1}
\Delta \Psi_{\mu }(x_{n}) \Psi_{\nu }(x_{n}) = \delta_{\mu \nu}.$$

The Sturm-Liouville problem (\ref{fdsch}), (\ref{fdbc}) can be rewritten
in a more visible form by using the symmetric tridiagonal (Jacobi) matrix
($N \times N$)
\begin{eqnarray*}
{\hat H} =  {\hat T} + {\hat J},
\end{eqnarray*}
\begin{eqnarray*}
{\hat T} = \left( \begin{array}{ccccccc} 2/\Delta ^{2} &
-1/\Delta ^2 & 0 & 0 & .  & .  & .\\ -1/\Delta ^2
& 2/\Delta ^{2} & -1/\Delta ^2 & 0
& .  & .  & . \\ 0 & -1/\Delta ^2 & 2/\Delta ^{2}
& -1/\Delta ^2 & . & . & .\\ .  & .  & . & . & . & . & .\\ .
& . & . & .  & . & . & .\\ . & . & . & . & 0 & -1/\Delta ^2
& 2/\Delta ^{2} \end{array} \right),
\end{eqnarray*}
\begin{equation}
{\hat J}= \left( \begin{array}{ccccccc}  V(x_{1}) &
u(x_{1}) & 0 & 0 & .  & .  & .\\ u(x_{1})
& V(x_{2}) & u(x_{2}) & 0
& .  & .  & . \\ 0 & u(x_{2}) & V(x_{3})
& u(x_{3}) & . & . & .\\ .  & .  & . & . & . & . & .\\ .
& . & . & .  & . & . & .\\ . & . & . & . & 0 & u(x_{N-1})
& V(x_{N}) \end{array} \right), \label{HJ}
\end{equation} which acts on the vector-column $\Psi _{\nu } \in {\mathbb
R}^{N}$:
\begin{eqnarray*} {\hat H} \Psi _{\nu }= E_{\nu } \Psi _{\nu }, \nonumber
\end{eqnarray*}
\begin{eqnarray*}
\Psi _{\nu }= \left( \begin{array}{c} \Psi (x_{1},E_{\nu }) \\
\Psi (x_{2},E_{\nu }) \\
. \\
. \\
. \\
\Psi (x_{N},E_{\nu })
\end{array} \right).
\end{eqnarray*}
This explicit form elucidates why the $u(x_{n})$ stands in front
of $\Psi (x_{n+1},E)$: that is because of the upper u-diagonal in
${\hat J}$ which is one element shorter than the main V-diagonal
and contains N-1 elements. The same coefficients form up the lower
diagonal (thanks to the matrix symmetry). Let us specially note
that  homogeneous boundary conditions generally different from
(\ref{fdbc}) would require some modification of the  matrix
representation (\ref{HJ}). In passing to the continuum limit
$\Delta \to 0$, it will be impossible to distinguish the u- and
V-diagonals, i.e., the resulting interaction will be simply the
sum of limiting values for u's and V's. A paragraph later, we
shall give the motivation for the appearance of the additional
diagonals in the interaction matrix ${\hat J}$.

Besides the energy levels, let us introduce additional fundamental
spectral parameters, namely, norming constants or spectral weight factors.
By the definition, these are the coefficients $c_{\nu }$ of
proportionality between the normalized eigenstates $\Psi_{\nu }(x_{n})$ and
the regular solutions $\varphi(x_{n},E_{\nu })$ at the eigenvalue energy,
$\varphi(0,E)=0, \varphi(x_{1},E)=\Delta$ [i.e. the derivative is equal to
1], $\varphi_{\nu }(x_{n}) \equiv \varphi(x_{n},E_{\nu}$):
\begin{eqnarray} \Psi_{\nu }(x_{n})=c_{\nu }\varphi_{\nu }(x_{n}).
\label{pphifd} \end{eqnarray} The continuum analog of the regular solution
satisfies $\varphi(0,E)=0, \varphi'(x,E)|_{x=0}=1$. The continuum
generalization of the spectral weight factors introduced is obvious.  In
classical inverse Sturm-Liouville problem (continuous coordinate), it is
well known that the double set of the spectral parameters $\{E_{\nu },
c_{\nu }\}$ uniquely specifies the potential.

Now we get to the core of our paper. Our task is to pose such a discrete
version of the inverse Sturm-Liouville problem that, in passing to the
continuum limit, as direct as possible reproduction  of all the result of
the continuum version is feasible.  So, the problem within which it
seems logical to work is posed as follows: Given the set $\{E_{\nu },
c_{\nu }\}$ with $c_{\nu }$ in (\ref{pphifd}),  the  potential matrix
${\hat J}$ with coefficients $V(x_{n})$ and $u(x_{n})$ is recovered
completely.  In principle, a question may arise whether the three-diagonal
Hamiltonian with the "non-local" u-coefficients is consistent with the
uniqueness of the potential recovery from the set $\{E_{\nu }, c_{\nu }\}$.
What does force the extra $u$'s?  As was already mentioned in the
introduction, any set $\{E_{\nu }, c_{\nu }\}$ can occur for a unique
three-diagonal ${\hat H}$.  Moreover, there is an extension of that result.
It is the theorem by Gladwell and Willms \cite{GladWil} in which the
statement was proved that a symmetric p-band matrix (a matrix with $2 p+1$
bands, p bands below the diagonal) may be uniquely constructed (apart from
certain sign ambiguities) from its eigenvalues and the first p components
of its normalized eigenvectors.  Hence, once we know all $E_{\nu }$'s and
the first eigenvector components, by virtue of Eq.  (\ref{pphifd}) these
are $\Delta c_{\nu }$ in our problem, we can uniquely restore 1-band, i.e.,
three-diagonal Hamiltonian (\ref{HJ}).  That also accounts for the u's.

Let us give an additional "half-heuristic" explanation of this fact.
If we have only a local potential $V(x_{n})$ with N values at
N points,  the number N of free parameters $\{V(x_{n})\}_{n=1}^{N}$
equals exactly the number of eigenvalues $E_{\nu }$. To the point, the
corresponding inverse problem has not a complete solution.
If we introduce both spectral parameters, $E_{\nu }$ and $c_{\nu }$,
while the single diagonal ${\hat J}$ persists, we shall really face a
problem of over-determination of the set $\{E_{\nu }, c_{\nu }\}$ that
contains  2 N-1 free parameters.  In fact, there are N levels
$E_{\nu}$ and N-1 parameters $c_{\nu}$ by virtue of the relation
$\sum_{\nu =1}^{N} c_{\nu }^{2} =1/\Delta^{3}$ that follows from
(\ref{pars1}) for $n=m=1$, while $V(x_{n})$  does only N ones.
It is introduction of  N-1 coefficients $u(x_{n}), n=1,...N-1$ into Eq.
(\ref{fdsch}) [or additional diagonals in (\ref{HJ})] that ensures the
equality of numbers of spectral data and interaction parameters.  In the
case of continuous coordinate, overfilling
the set of spectral parameters reveals itself only in many-dimensional $D
\ge 2$ problems, so we shall manage to restore one-dimensional local
potential by the complete spectral set [see (\ref{vK})].

The functions $\varphi_{\nu }(x_{n})$ can be considered as vectors in
special Hilbert (Euclidean, to be precise) space, in which the
coordinate $x_{n} $ numbers the eigenvectors and energy index $\nu$ is only
used to denote the $\nu $th vector component. The inner product in that
space is determined by the measure given by the spectral weight factors
$c_{\nu}$.  In fact, the Parseval's completeness relation
\begin{eqnarray} \sum_{\nu =1}^{N} \Psi _{\nu }(x_{m})
\Psi _{\nu }(x_{n}) = \delta_{mn}/\Delta \label{Pars} \end{eqnarray} can
be rewritten  using Eq.(\ref{pphifd}) as
\begin{eqnarray} \sum_{\nu
=1}^{N} c_{\nu }^{2} \varphi_{\nu }(x_{m}) \varphi_{\nu
}(x_{n})=\delta_{mn}/\Delta.  \label{pars1} \end{eqnarray}
Let us consider this expression as an orthogonality relation
for the vectors $\varphi_{\nu }(x_{m})$ and $\varphi_{\nu }(x_{n})$ (in the
limit $\Delta \to 0$, the  "numbers" $x_{m,n}$ of the vectors
become continuous variable $x$).  Here, the inner product
is given by not simply a sum over energy index $\nu $ but a sum with a
weight (measure) $c_{\nu }^2$.

Different potentials correspond to different weight factors determining
the metrics of our "energy space" but the relation (\ref{pars1})
holds true for any potential.  In the classical variant, the inverse
problem can be treated as a transition to the sought potential
$\stackrel{\circ}{V}(x_{n}) \to V(x_{n})$ from a certain "initial" (in what
follows we shall use the symbol "$\circ$" to denote everything related
to the initial system) potential $\stackrel{\circ}{V}(x_{n})$, for
which all the solutions $\stackrel{\circ}{\varphi}_{\nu }(x_{n})$ and the
whole spectral set
$\{\stackrel{\circ}{E}_{\nu }, \stackrel{\circ}{c}_{\nu }\}$ are known, and
the relation (\ref{pars1}) is valid:  \begin{eqnarray} \sum_{\nu =1}^{N}
\stackrel{\circ}{c}_{\nu }^{2} \stackrel{\circ}{\varphi}_{\nu }(x_{m})
\stackrel{\circ}{\varphi}_{\nu }(x_{n})=\delta_{mn}/\Delta.  \label{pars2}
\end{eqnarray}

All this gives us a hint for deriving new solutions corresponding to
the given spectral set $\{E_{\nu }, c_{\nu }\}_{\nu =1}^{N}$. Although
we do not know yet the sought potential matrix ${\hat J}$, we beforehand
know that the regular solutions $\varphi_{\nu}(x_{n})$ to these potentials
must satisfy the orthogonality relation (\ref{pars1}) with the new
$c_{\nu }$. We shall catch at this fact and use the orthogonality
relation as a central criterion in finding new eigenvectors (solutions
$\varphi_{\nu}(x_{n})$). Changing the metrics of Euclidean
space in replacing $\stackrel{\circ}{c}_{\nu} \to c_{\nu}$ results in that
the "old" unit vectors $\stackrel{\circ}{\varphi}_{\nu }(x_{n})$ are no
longer orthogonal. So the idea is as follows: Once new unit vectors must
satisfy Eq.(\ref{pars1}), we could obtain them, e.g., orthogonalizing
the $\stackrel{\circ}{\varphi}_{\nu }(x_{n})$ with the new weight $c_{\nu
}^2$ by the Gram-Schmidt scheme.
In other words, the new vectors obtained by that way and satisfying
(\ref{pars1}) with the weight multipliers  $c_{\nu }^2$ will be the
solutions to the new potentials $V$'s and $u$'s.  Indubitably, this makes
sense only when the procedure really gives the desired vectors, i.e., it is
unique (see the proof further on).

For simplicity, we shall at first think $\stackrel{\circ}{E}_{\nu
} = E_{\nu }$.  Let us recall this standard orthogonalization
procedure for the example of two initially non-orthogonal (in
sense of new weight function) vectors (i.e. when N=2)
$\stackrel{\circ}{\varphi }_{\nu }(x_{1}),
\stackrel{\circ}{\varphi }_{\nu }(x_{2}), \nu = 1,2.$ As a first
unit vector $\varphi_{\nu }(x_{1})$ of the new system, we take the
unchanged unit vector $\stackrel{\circ}{\varphi }_{\nu }(x_{1})$,
and the second unit vector is constructed from the second
unaltered one, only we have to subtract everything superfluous
(parallel to $\stackrel{\circ}{\varphi }_{\nu }(x_{1})$), for the
orthogonality with the new measure:  $$\varphi_{\nu
}(x_{1})=\stackrel{\circ}{\varphi }_{\nu }(x_{1}); \,\,\,
\varphi_{\nu }(x_{2})=\stackrel{\circ}{\varphi }_{\nu }(x_{2})+
\Delta K(x_{2},x_{1})\stackrel{\circ}{\varphi }_{\nu }(x_{1}).$$
The coefficient $K(x_{2},x_{1})$ is derived from the condition of
orthogonality of the new vectors with the new weight $c_{\nu }$:
$$\varphi _{\nu }(x_{2}) \perp \varphi _{\nu }(x_{1}) \equiv \stackrel{\circ
}{\varphi }_{\nu }(x_{1}).$$
We have
\begin{eqnarray} \sum_{\nu =1}^{2}
c_{\nu }^{2} \stackrel{\circ}{\varphi }_{\nu } (x_{1}) \stackrel{\circ
}{\varphi }_{\nu }(x_{2}) + \Delta \sum_{\nu =1}^{2} c_{\nu }^{2}
K(x_{2},x_{1}) \stackrel{\circ }{\varphi }_{\nu }^{2}(x_{1})   = 0
\Longrightarrow \nonumber \\
\hspace*{-1cm} K(x_{2},x_{1}) + \sum_{\nu=1}^{2} c_{\nu}^{2}
\stackrel{\circ}{\varphi }_{\nu }(x_{1}) \stackrel{\circ }{\varphi
}_{\nu }(x_{2})  + \Delta K(x_{2},x_{1}) \sum_{\nu =1}^{2} (c_{\nu
}^{2} - \stackrel{\circ}{c}_{\nu}^2) \stackrel{\circ }{\varphi
}_{\nu }^{2}(x_{1}) = 0, \label{GL2i} \end{eqnarray} where we add
and subtract the  term $\stackrel{\circ}{c}_{\nu}^2$ from the
multiplier  $c_{\nu }^2$  and, furthermore, use Eq. (\ref{pars2}).
We can rewrite the last equality in the form as follows (extremely
simplified two-dimensional "prototype" of the inverse problem
equation):
\begin{eqnarray} K(x_{2},x_{1}) + Q(x_{2},x_{1}) + \Delta K(x_{2},x_{1})
Q(x_{1},x_{1}) = 0, \label{GL2} \end{eqnarray} where \begin{eqnarray}
Q(x_{m},x_{n}) = \sum_{\nu =1}^{N=2} c_{\nu }^{2} \stackrel{\circ}{\varphi
}_{\nu }(x_{m}) \stackrel{\circ}{\varphi }_{\nu }(x_{n}) - \nonumber \\
\sum_{\mu =1}^{N=2}\stackrel{\circ}{c}_{\mu }^{2} \stackrel{\circ}{\varphi
 }_{\mu }(x_{m}) \stackrel{\circ}{\varphi }_{\mu }(x_{n}); \quad m,n=1,2.
\label{Q2} \end{eqnarray}

In general case of  N-dimensional Euclidean space we shall follow the same
scheme. In doing so, it is possible to take into account  the case
when the levels change: $\stackrel{\circ}{E}_{\nu } \ne E_{\nu }$.  We must
orthogonalize  N vectors by the measure $c_{\nu}^2$:
$\stackrel{\circ}{\varphi }(x_{m},E_{\nu })$, $(m=1,2,...N, \nu
=1,2,...N)$.  Consequently, we have for the new solutions
\begin{eqnarray} \varphi (x_{m},E_{\nu})= \stackrel{\circ}{\varphi
}(x_{m},E_{\nu}) + \sum_{n=1}^{m-1} \Delta K(x_{m},x_{n})
\stackrel{\circ}{\varphi }(x_{n},E_{\nu}), \label{fKfnm} \end{eqnarray}
where the coefficients K [the kernel of the transformation operator
(\ref{fKfnm})] follow from the conditions of the orthogonality of
new vectors  (by measure $c_{\nu}^2$) $\varphi (x_{m},E_{\nu }) \,
(m=1,2,...N)$:  $$\varphi (x_{m
> n},E_{\nu}) \perp \varphi (x_{n},E_{\nu}),$$ which lead
to the system  of algebraic equations for K -- discrete analog
of central equations of the inverse problem:
\begin{eqnarray} K(x_{m},x_{n})+Q(x_{m},x_{n})+ \sum_{p=1}^{m-1} \Delta
K(x_{m},x_{p}) Q(x_{p},x_{n})=0, \quad m>n, \label{GLnm}
\end{eqnarray}
where $Q(x_{n},x_{m})$ is determined as in (\ref{Q2}), only the
values $m$ and $n$ are no longer restricted by 1 and 2, and the indices
$\mu$ and $\nu$ number solutions at initial and shifted energy levels,
respectively.
\begin{eqnarray}
Q(x_{m},x_{n}) = \sum_{\nu =1}^{N} c_{\nu }^{2} \stackrel{\circ}{\varphi
}(x_{m},E_{\nu}) \stackrel{\circ}{\varphi }(x_{n},E_{\nu})  \nonumber \\
- \sum_{\mu =1}^{N}\stackrel{\circ}{c}_{\mu }^{2}
\stackrel{\circ}{\varphi }(x_{m},\stackrel{\circ}{E}_{\mu})
\stackrel{\circ}{\varphi }(x_{n}, \stackrel{\circ}{E}_{\mu}).
\label{QN} \end{eqnarray} Let us note that the form (\ref{fKfnm})
ensures the desired boundary condition for the regular solution:
$\varphi (x_{0},E_{\nu})=0; \varphi (x_{1},E_{\nu})=\Delta$. The
system Eq. (\ref{GLnm}) of the recurrence computation of the K's
provides them uniquely. We can formally introduce the diagonal
terms $K(x_{n},x_{n})$ (bearing no relation to $\varphi(x,E_{\nu
})$) such as $K(x_{n+1},x_{n})-K(x_{n},x_{n}) \sim O(\Delta)$
which will be useful in what follows.

When the first $m+1$ unit vectors $\stackrel{\circ}{\varphi
}(x_{i},E_{\nu}), \enskip i=1,..,m+1$ are orthogonalized, this
corresponds to an intermediate submatrix-block  transformation of
the initial Jacobi-like operator (\ref{HJ}) so that
\begin{eqnarray}
{\hat J}= \left( \begin{array}{cc} {\hat J}_{m} & {\bf 0} \\
{\bf 0} & \hat {\stackrel{\circ}{J}}_{N-m} \end{array} \right),
\nonumber \\
{\hat J}_{m}= \left( \begin{array}{ccccccc} V(x_{1}) & u(x_{1}) &
0 & 0 & .  & .  & .\\ u(x_{1}) & V(x_{2}) & u(x_{2})
& 0 & .  & . & . \\ 0 & u(x_{2}) & V(x_{3}) & u(x_{3}) & . & . & .\\
.  & .  & . & . & . & . & .\\ . & . & . & .  & . & . & .\\ . & . &
. & 0 & u(x_{m-1})  & V(x_{m}) & \stackrel{\circ}{u}(x_{m})
\end{array} \right), \nonumber \\
\hat {\stackrel{\circ}{J}}_{N-m}= \left(
\begin{array}{ccccccc} \stackrel{\circ}{u}(x_{m}) &
\stackrel{\circ}{V}(x_{m+1})
& \stackrel{\circ}{u}(x_{m+1}) & 0 & .  & .  & .\\
0 & \stackrel{\circ}{u}(x_{m+1}) & \stackrel{\circ}{V}(x_{m+2}) &
\stackrel{\circ}{u}(x_{m+2}) & 0 & .  & .    \\
.  & .  & . & . & . & . & .\\ .  & . & . & .  & . & . & .\\ . & .
& . & . & 0 & \stackrel{\circ}{u}(x_{N-1}) &
\stackrel{\circ}{V}(x_{N}) \end{array} \right),
 \label{mJ}
\end{eqnarray}
where the symbols ${\bf 0}$ in the top-right and bottom-left
corners of the ${\hat J}$-matrix denote zero $(m \times N-m-1)$
and $(N-m \times m-1)$ matrices, respectively. The submatrix
${\hat J}_{m}$ is formed up by the perturbed coefficients while
the $\hat {\stackrel{\circ}{J}}_{N-m}$ is not affected yet by the
transformation associated with the reorthogonalization. Note that
the last row of the submatrix ${\hat J}_{m}$ contains only two
transformed elements, $u(x_{m-1})$ and $V(x_{m})$, apart from the
$\stackrel{\circ}{u}(x_{m})$. This is because  the $(m \times m)$
{\it quadratic} submatrix was transformed only which contains, in
its last row,  two non-zero elements mentioned. The coefficients
of that intermediate transformation block may be found from
formulas (\ref{vfd1}) and (\ref{vfd2}) where one should substitute
$\stackrel{\circ}{u}(x_{m})$ for $u(x_{m})$. See also the formulas
(\ref{vfd1}) and (\ref{vfd2}) and the subsequent discussion.

Now we are ready to give the proof that the above procedure is
unique. Let us carry out it by induction. Suppose that the desired
transformed vectors are uniquely given by the equation
(\ref{fKfnm}) for $m \le {\bar N}, \enskip {\bar N} < N$ for a
certain ${\bar N}$ being the integer. For ${\bar N}=1$, this is
verified trivially.  Let us show that the formula (\ref{fKfnm})
holds true at the point $x_{{\bar N}+1}$. Indeed, the $\varphi
(x_{{\bar N}+1},E_{\nu})$, being orthogonal to all $\varphi
(x_{m},E_{\nu}), m \le {\bar N}$, can be sought, in principle, as
a combination of the initial $\stackrel{\circ}{\varphi
}(x_{n},E_{\nu})$ for all $n$. The coefficients of such a
hypothetical combination (there are N pieces of them in all) have
to be determined from the condition of orthogonality of $\varphi
(x_{{\bar N}+1},E_{\nu})$ to both $\varphi (x_{m},E_{\nu}),
m=1,... {\bar N}$ and certain $N-{\bar N}$ unknown vectors from
the new orthogonal set (\ref{pars1}). For ascertaining what we
shall do now, we involve  the Schr\"odinger equation
(\ref{fdsch}), which is a recurrence procedure of the step-by-step
computation of the $\varphi (x_{n},E_{\nu})$. Specifically, we
mean the block operator ${\hat J}_{{\bar N}}$ in (\ref{mJ}), last
row:
\begin{eqnarray*}
-\frac{\varphi(x_{{\bar N}+1},E_{\nu})-2 \varphi(x_{{\bar
N}},E_{\nu}) +\varphi(x_{{\bar N}-1},E_{\nu})}{\Delta ^2}+
V(x_{{\bar N}}) \varphi(x_{{\bar N}},E_{\nu})  \nonumber \\
+ \stackrel{\circ}{u}(x_{{\bar N}}) \varphi(x_{{\bar
N}+1},E_{\nu}) + u(x_{{\bar N}-1}) \varphi(x_{{\bar N}-1},E_{\nu})
= E_{\nu} \varphi(x_{{\bar N}},E_{\nu}).
\end{eqnarray*}
It is seen that $\stackrel{\circ}{u}(x_{{\bar N}}) \varphi
(x_{{\bar N}+1},E_{\nu})$ is a linear combination of $u(x_{{\bar
N}-1}) \varphi (x_{{\bar N}-1},E_{\nu})$, $V(x_{{\bar N}}) \varphi
(x_{{\bar N}},E_{\nu})$ and $E_{\nu } \varphi (x_{{\bar
N}},E_{\nu})$. Consequently, by the assumptions of the validity of
(\ref{fKfnm}) for $m \le {\bar N}$, $\stackrel{\circ}{u}(x_{{\bar
N}}) \varphi (x_{{\bar N}+1},E_{\nu})$ can only be represented
through $V(x_{{\bar N}}) K(x_{{\bar N}},
x_{m})\stackrel{\circ}{\varphi }(x_{m},E_{\nu}), \enskip m=1,...
{\bar N}-1$; $V(x_{{\bar N}}) \stackrel{\circ}{\varphi }(x_{{\bar
N}},E_{\nu})$; $u(x_{{\bar N}-1}) K(x_{{\bar N-1}},
x_{m})\stackrel{\circ}{\varphi }(x_{m},E_{\nu}), \enskip m=1,...
{\bar N}-2$; $u(x_{{\bar N}-1}) \stackrel{\circ}{\varphi
}(x_{{\bar N}-1},E_{\nu})$; $K(x_{{\bar N-1}}, x_{m}) E_{\nu
}\stackrel{\circ}{\varphi }(x_{m},E_{\nu}), \enskip m=1,... {\bar
N}-1$; and $E_{\nu }\stackrel{\circ}{\varphi }(x_{{\bar
N}},E_{\nu})$. For the last term, we find from the non-perturbed
Schr\"odinger equation  that $E_{\nu }\stackrel{\circ}{\varphi
}(x_{{\bar N}}, E_{\nu})$ is expressed through
$\stackrel{\circ}{u}(x_{{\bar N}}) \stackrel{\circ}{\varphi
}(x_{{\bar N + 1}}, E_{\nu})$, $\stackrel{\circ}{V}(x_{{\bar N}})
\stackrel{\circ}{\varphi }(x_{{\bar N}}, E_{\nu})$,
$\stackrel{\circ}{u}(x_{{\bar N}-1}) \stackrel{\circ}{\varphi
}(x_{{\bar N - 1}}, E_{\nu})$ [and similarly for other $E_{\nu
}\stackrel{\circ}{\varphi }(x_{m},E_{\nu})$] and, finally, the
$\varphi (x_{{\bar N}+1},E_{\nu})$ must be sought as a linear
combination of $\stackrel{\circ}{\varphi }(x_{m}, E_{\nu}),
\enskip m=1,... {\bar N}+1$. In other words, $\varphi (x_{{\bar
N}+1},E_{\nu}) \in$ span$\{\stackrel{\circ}{\varphi
}(x_{m},E_{\nu}) \}_{m=1}^{{\bar N}+1}$, besides that $\varphi
(x_{{\bar N}+1},E_{\nu}) \perp$ span$\{\stackrel{\circ}{\varphi
}(x_{m}, E_{\nu}) \}_{m=1}^{{\bar N}}=$ span$\{\varphi (x_{m},
E_{\nu}) \}_{m=1}^{{\bar N}}$. It is well known that the
Gram-Schmidt orthogonalization enables a unique solution
satisfying these two conditions. Note that $({\bar N}+1)$th term
$\stackrel{\circ}{u}(x_{{\bar N}}) \varphi (x_{{\bar
N}+1},E_{\nu})$ is a combination of the summand
$\stackrel{\circ}{u}(x_{{\bar N}}) \stackrel{\circ}{\varphi }
(x_{{\bar N}+1},E_{\nu})$ (with the same coefficient) and other
terms with $m \le {\bar N}$. In other words, in the decomposition
of $\varphi (x_{{\bar N}+1},E_{\nu})$, we have the term
$\stackrel{\circ}{\varphi } (x_{{\bar N}+1},E_{\nu})$ with the
unit coefficient. Consequently, $\varphi (x_{{\bar N}+1},E_{\nu})$
is represented in the form (\ref{fKfnm}) again (QED).

It should be noted that the formula (\ref{fKfnm}) is also valid
for solutions at energies $E$ lying between the levels $E_{\nu }$
where the regular solutions, being the Cauchy problem solutions,
are defined (though non-physical). Do not confuse "running" energy
values at which the solutions $\varphi(x_{m},E)$ are defined with
the energies occurring in the inverse problem equations
(\ref{QN}). Indeed, let us decompose $\varphi (x_{m},E)$ into the
complete set of the solutions $\varphi (x_{m},E_{\nu})$ (in sense
of the usual inner product $\sum_{m=1}^{N} \Delta c_{\nu} c_{\mu
}\varphi (x_{m},E_{\nu }) \varphi (x_{m},E_{\mu })$):
\begin{eqnarray}
\varphi (x_{m},E)=\sum_{\nu=1}^{N}\xi(E,E_{\nu})\varphi
(x_{m},E_{\nu}); \enskip \xi(E,E_{\nu})=\sum_{m=1}^{N} \Delta
c_{\nu} \varphi (x_{m},E_{\nu }) \varphi (x_{m},E). \label{dec1}
\end{eqnarray}
Since  the $\varphi (x_{m},E_{\nu})$'s are expressed, in
accordance to (\ref{fKfnm}), through the unperturbed
$\stackrel{\circ}{\varphi }(x_{m},E_{\nu})$ then we shall expand
them, too, in a complete set of the old solutions
$\stackrel{\circ}{\varphi }(x_{m},\stackrel{\circ}{E}_{\mu})$:
\begin{eqnarray}
\stackrel{\circ}{\varphi}(x_{m},E_{\nu
})=\sum_{\mu=1}^{N}\zeta(E_{\nu},\stackrel{\circ}{E}_{\mu})\stackrel{\circ}{\varphi}
(x_{m},\stackrel{\circ}{E}_{\mu}); \enskip
\zeta(E_{\nu},\stackrel{\circ}{E}_{\mu})=\sum_{m=1}^{N} \Delta
\stackrel{\circ}{c}_{\mu} \stackrel{\circ}{\varphi}(x_{m},E_{\nu
}) \stackrel{\circ}{\varphi} (x_{m},\stackrel{\circ}{E}_{\mu}).
\nonumber \\ \label{dec2}
\end{eqnarray}

Combining (\ref{fKfnm}), (\ref{dec1}) and (\ref{dec2}), we get the
following expression for the new solutions at arbitrary $E$:
\begin{eqnarray}  \varphi (x_{m},E)= \sum_{\mu, \nu}
A(E,E_{\nu },\stackrel{\circ}{E}_{\mu}) \stackrel{\circ}{\varphi
}(x_{m},\stackrel{\circ}{E}_{\mu})  + \sum_{\mu, \nu}
\sum_{n=1}^{m-1} \Delta K(x_{m},x_{n}) \nonumber \\ \times
A(E,E_{\nu },\stackrel{\circ}{E}_{\mu}) \stackrel{\circ}{\varphi
}(x_{n},\stackrel{\circ}{E}_{\mu}); \enskip A(E,E_{\nu
},\stackrel{\circ}{E}_{\mu})=\xi(E,E_{\nu})
\zeta(E_{\nu},\stackrel{\circ}{E}_{\mu}). \label{fearb}
\end{eqnarray} In the limit when the new and old spectral
parameters coincide, $K$ vanishes and, hence, $\varphi(x_{m},E)$
turn into unperturbed solution
$$\stackrel{\circ}{\varphi}(x_{m},E)= \sum_{\mu, \nu } A(E,E_{\nu },\stackrel{\circ}{E}_{\mu})
\stackrel{\circ}{\varphi } (x_{m},\stackrel{\circ}{E}_{\mu}).$$
Substituting this expression in  (\ref{fearb}), we get:
\begin{eqnarray} \varphi (x_{m},E)= \stackrel{\circ}{\varphi }(x_{m},E) +
\sum_{n=1}^{m-1} \Delta K(x_{m},x_{n}) \stackrel{\circ}{\varphi }(x_{n},E).
\label{fearb1} \end{eqnarray}
Let us stress here that $K$ is independent of energy $E$.
The formulas (\ref{fearb1}), (\ref{GLnm}) and (\ref{QN})
give the expression for $K$ in the form of sum of
products of the old solutions and transformed ones:
\begin{eqnarray}
K(x_{m},x_{n}) = -\sum_{\nu }^{N} c_{\nu}^{2}\varphi
(x_{m},E_{\nu}) \stackrel{\circ}{\varphi }(x_{n},E_{\nu}) +
\sum_{\mu }^{N}\stackrel{\circ}{c}^{2}_{\mu } \varphi
(x_{m},\stackrel{\circ}{E}_{\mu}) \stackrel{\circ}{\varphi
}(x_{n},\stackrel{\circ}{E}_{\mu}). \nonumber \\
\label{kp0pnm} \end{eqnarray}

It remains to obtain equations for the transformed potentials $V$
and $u$. We already know the solutions of Eq. (\ref{fdsch}) with
the unknown potentials $V(x_{n})$ and $u(x_{n})$ [see formulas
(\ref{fKfnm}) and (\ref{GLnm})], i.e. eigenvectors of the new
Hamiltonian plus associated eigenvalues $E_{\nu }$. As was
mentioned, by virtue of the theorem by Gladwell and Willms
\cite{GladWil},  that is enough for the three-diagonal Hamiltonian
matrix (with off-diagonal elements) to be uniquely recovered.
These authors used the block Lanczos algorithm. However, we shall
apply an outwardly different method pursuing the aim of
reproducing final formulas in the continuum limit.  Let us
multiply both parts of the Schr\"odinger equation (\ref{fdsch})
for the solutions $\varphi (x_{m},\stackrel{\circ}{E}_{\mu})$ [Eq.
(\ref{fearb1})] and $\stackrel{\circ}{\varphi
}(x_{n},\stackrel{\circ}{E}_{\mu})$ by $\stackrel{\circ}{\varphi
}(x_{n}, \stackrel{\circ}{E}_{\mu})$ and
$\varphi(x_{m},\stackrel{\circ}{E}_{\mu})$, respectively, sum over
$\mu $ with weight $\stackrel{\circ}{c}_{\mu }^{2}$ and subtract
from each other the resulting expressions. At fixed $m$, we
perform this procedure for $n=m, m-1,...$.  In calculating sums
(over $\mu$) one should take into account the relation
(\ref{pars2}).  As a result, we get to the following equations for
$V$ and $u$:  \begin{eqnarray} \{V(x_{m}) -
\stackrel{\circ}{V}(x_{n})\} K(x_{m},x_{n})+ u(x_{m})
K(x_{m+1},x_{n}) - \stackrel{\circ}{u}(x_{n}) \nonumber \\ \times
K(x_{m},x_{n+1}) + u(x_{m-1})
K(x_{m-1},x_{n})-\stackrel{\circ}{u}( x_{n-1}) K(x_{m},x_{n-1})
\nonumber \\ = \frac{K(x_{m+1},x_{n})-2
K(x_{m},x_{n})+K(x_{m-1},x_{n})}{\Delta^2} \nonumber \\ -
\frac{K(x_{m},x_{n+1})-2
K(x_{m},x_{n})+K(x_{m},x_{n-1})}{\Delta^2}, \enskip n \le m-2;
\label{vfd1} \end{eqnarray} and for $n=m,m-1$
\begin{eqnarray} \left\{ \begin{array}{l}
\frac{u(x_{m-1})-\stackrel{\circ}{u}(x_{m-1})}{\Delta} =
\frac{K(x_{m+1},x_{m-1})-K(x_{m},x_{m-2})}{\Delta ^2} \\
- V(x_{m}) K(x_{m},x_{m-1})+\stackrel{\circ}{V}(x_{m-1}) K(x_{m},x_{m-1}) \\
- u(x_{m}) K(x_{m+1},x_{m-1})+\stackrel{\circ}{u}(x_{m-2}) K(x_{m},x_{m-2}),
\enskip n=m-1 \\
\frac{V(x_{m})-\stackrel{\circ}{V}(x_{m})}{\Delta}=
\frac{K(x_{m+1},x_{m})-K(x_{m},x_{m-1})}{\Delta ^2} \\
- u(x_{m}) K(x_{m+1},x_{m}) + K(x_{m},x_{m-1}) \stackrel{\circ}{u}
(x_{m-1}), \enskip n=m,
\end{array}
\right.
\label{vfd2}
\end{eqnarray}
where the terms $K(x_{m},x_{n})$ for which $m,n >N$ or $m,n <1$
are omitted. But for $n=m+1$ we obtain that
$u(x_{m})=\stackrel{\circ}{u}(x_{m})$.  There is nothing strange
in it because summation is carried out for the term
$\stackrel{\circ}{\varphi } (x_{m+1},\stackrel{\circ}{E}_{\mu})$
that is {\it orthogonal} to all $\stackrel{\circ}{\varphi}
(x_{n},\stackrel{\circ}{E}_{\mu})$, $n < m+1$. But the kernel $K$
containing all the information about new solutions just stands at
these summands, see formula (\ref{fearb1}) for $n=m+1$. In other
words, for the case $n=m+1$ the summation expunges everything that
bears a relation to the new system under construction. Indeed,
from the Schr\"odinger equation for $\stackrel{\circ}{\varphi
}(x_{m+1}, \stackrel{\circ}{E}_{\mu})$ multiplied by $\varphi
(x_{m},\stackrel{\circ}{E}_{\mu})$ and summed over $\mu$ with the
weight $\stackrel{\circ}{c}_{\mu }^{2}$ we have
$$\sum_{\mu =1}^{N}
\stackrel{\circ}{c}_{\mu }^{2} \stackrel{\circ}{E}_{\mu} \varphi
(x_{m},\stackrel{\circ}{E}_{\mu}) \stackrel{\circ}{\varphi
}(x_{m+1},
\stackrel{\circ}{E}_{\mu})=\frac{\stackrel{\circ}{u}(x_{m})}{\Delta}
- \frac{1}{\Delta ^3}.$$ How should one treat this? One variant is
that $u(x_{m})=\stackrel{\circ}{u}(x_{m})$, which corresponds to
restoration of a  $m \times m$ submatrix for which the element
$\stackrel{\circ}{u}(x_{m})$ is an outer one, see (\ref{mJ}). This
can also serve as a proof of the (\ref{mJ}). Second interpretation
is that $u(x_{m})\ne \stackrel{\circ}{u}(x_{m})$ (but not for
$m=N$), nevertheless (i.e, the above procedure does not work).
This takes the place when we have the whole ${\hat J}$-matrix
transformed but the equations (\ref{vfd2}) by themselves did not
allow the computation of $u(x_{m})$. How then  to uniquely restore
the ${\hat J}$ will be discussed a bit later, but now we must
ascertain for ourselves how it is possible that the same solution
$\varphi (x_{m},E)$ may satisfy the Schr\"odinger equation
(\ref{fdsch}) with {\it different} potential coefficients
$\stackrel{\circ}{u}(x_{m})$ and $u(x_{m})$ at $x_{m+1}$. The
matter is that we deal with (finite-difference) non-local
potential and this ambiguity is just characteristic of it. Indeed,
let $\varphi (x_{m},E)$ satisfy the Schr\"odinger equations with
both $\{u_{1}(x_{m-1}), V_{1}(x_{m}), u_{1}(x_{m})\}$ and
$\{u_{2}(x_{m-1}), V_{2}(x_{m}), u_{2}(x_{m})\}$. Subtracting
these equations from each other we have
\begin{eqnarray*}
[V_{1}(x_{m})- V_{2}(x_{m})]\varphi (x_{m},E) +
[u_{1}(x_{m})-u_{2}(x_{m})] \varphi (x_{m+1},E) \\ +
[u_{1}(x_{m-1})-u_{2}(x_{m-1})]\varphi (x_{m-1},E)=0.
\end{eqnarray*}
If we have only one (local) potential coefficient then it would be
the same. But now, for several non-local potential coefficients
coupling neighbour $x$-points, that equation clearly demonstrates
that $V_{1}(x_{m})- V_{2}(x_{m})$ and others may all be non-zero.
Summing up this discussion, we have elucidated that the procedure
used for derivation of (\ref{vfd1}) and (\ref{vfd2}) cannot
distinguish all the variants of $u(x_{m})$-coefficient
determination proceeding from a general incapability of giving a
unique non-local interaction  associated with a certain solution
of Schr\"odinger equation.

However, for the whole vector $\varphi (x_{m},E_{\nu })$, i.e. the
solution defined at {\it all} the points $x_{m}, \enskip
m=1,...,N$ we are able to uniquely derived the quadratic potential
matrix ${\hat J}$ whose eigenvectors are $\varphi (x_{m},E_{\nu
})$. Taking $m=N$ we first find $u(x_{N-1})$, $V(x_{N})$ and
$u(x_{N})$. Of course, this requires the knowledge of $u(x_{N})$.
But we have no more equations for determining the potential
coefficient $u(x_{N})$. However, we see that the $u(x_{N})$ is a
continuation of the last $N$th row of the matrix ${\hat J}$. This
resembles the case with the unfinished restoration of the ${\hat
J}$, see (\ref{mJ}), i.e. the potential perturbation (in form of a
quadratic matrix) never reached the $u(x_{N})$. That is, the
$u(x_{N})$ is independent of the transformation generated by
$K$-coefficients. But then, taking $\{E_{\nu},
c_{\nu}\}=\{\stackrel{\circ}{E}_{\nu},
\stackrel{\circ}{c}_{\nu}\}$, we see that $K=0$ and $u(x_{N})$
exactly corresponds to the reference potential. Thus, we have
$u(x_{N})=\stackrel{\circ}{u}(x_{N})$. Next, at the point $x_{N}$
we have, instead of  (\ref{vfd2}), the following system of
equations
\begin{eqnarray} \left\{ \begin{array}{l}
\frac{u(x_{N-1})-\stackrel{\circ}{u}(x_{N-1})}{\Delta} =
\frac{\sum_{\mu =1}^{N} \stackrel{\circ}{c}_{\mu }^{2} \varphi
(x_{N+1},\stackrel{\circ}{E}_{\mu}) \stackrel{\circ}{\varphi
}(x_{N-1},\stackrel{\circ}{E}_{\mu})-K(x_{N},x_{N-2})}{\Delta ^2} \\
- V(x_{N}) K(x_{N},x_{N-1})+\stackrel{\circ}{V}(x_{N-1}) K(x_{N},x_{N-1}) \\
- u(x_{N}) \sum_{\mu =1}^{N} \stackrel{\circ}{c}_{\mu }^{2}
\varphi (x_{N+1},\stackrel{\circ}{E}_{\mu})
\stackrel{\circ}{\varphi
}(x_{N-1},\stackrel{\circ}{E}_{\mu})+\stackrel{\circ}{u}(x_{N-2})
K(x_{N},x_{N-2}), \\
\frac{V(x_{N})-\stackrel{\circ}{V}(x_{N})}{\Delta}=
\frac{\sum_{\mu =1}^{N} \stackrel{\circ}{c}_{\mu }^{2} \varphi
(x_{N+1},\stackrel{\circ}{E}_{\mu}) \stackrel{\circ}{\varphi
}(x_{N},\stackrel{\circ}{E}_{\mu})-K(x_{N},x_{N-1})}{\Delta ^2} \\
- u(x_{N}) \sum_{\mu =1}^{N} \stackrel{\circ}{c}_{\mu }^{2}
\varphi (x_{N+1},\stackrel{\circ}{E}_{\mu})
\stackrel{\circ}{\varphi }(x_{N},\stackrel{\circ}{E}_{\mu}) +
K(x_{N},x_{N-1}) \stackrel{\circ}{u} (x_{N-1}),
\end{array}
\right. \nonumber \\ \label{vfdN}
\end{eqnarray}
where $\varphi (x_{N+1},\stackrel{\circ}{E}_{\mu})$ is found from
the Schr\"odinger equation (\ref{fdsch}):
\begin{eqnarray}
\varphi (x_{N+1},\stackrel{\circ}{E}_{\mu})=\frac{\Delta
^2}{1-\Delta ^2 u(x_{N})} [u(x_{N-1}) \varphi
(x_{N-1},\stackrel{\circ}{E}_{\mu}) \nonumber
\\ + V(x_{n}) \varphi
(x_{N},\stackrel{\circ}{E}_{\mu})-  \stackrel{\circ}{E}_{\mu }
\varphi (x_{N},\stackrel{\circ}{E}_{\mu})] -\frac{- 2 \varphi
(x_{N},\stackrel{\circ}{E}_{\mu}) + \varphi
(x_{N-1},\stackrel{\circ}{E}_{\mu})}{1-\Delta ^2 u(x_{N})},
\label{uN}
\end{eqnarray}
$\varphi (x_{n},\stackrel{\circ}{E}_{\mu})$ given by
(\ref{fearb1}) and $u(x_{N})=\stackrel{\circ}{u}(x_{N})$. From
(\ref{vfdN}) and (\ref{uN}) we obtain $V(x_{N})$ and $u(x_{N-1})$.
We then substitute the value $u(x_{N-1})$ (by virtue of the
symmetry of potential matrix) into equations (\ref{vfd1}) for
$m=N-1; n=m-2, m-3$ from which we find, in turn, $V(x_{N-1})$ and
$u(x_{N-2})$. Afterward, we substitute this last coefficient into
equation (\ref{vfd1}) for $m=N-2; n=m-2, m-3$ and get $V(x_{N-2})$
and $u(x_{N-3})$ and so on. Thus these equations allow the
computation of V and u via the solutions of the inverse problem
equation (\ref{GLnm}) -- the coefficients $K(x_{m}, x_{n}), m>n$
(plus additional requirement concerning $u(x_{N})$). For any
finite N, these linear equations are uniquely solved. But, with
the N large, the numerical instability increases that leads to the
well known problem of the ill-posed inversion procedure in the
continuum limit. However, let us drop the discussion on that
problem here, especially as we only want to reproduce the
expression for the continuous potential.  So we shall keep on
dealing with equations (\ref{vfd1}) and (\ref{vfd2}) and next show
that passing to the continuum limit in these equations will lead
us to the classical results of the Sturm-Liouville inverse
problem.

\section{Continuum limit}

Let us now pass to the limit of the continuous variable $x$, i.e.
to the limit $\Delta \to 0$ ($N \to \infty$) so that $\Delta N =
\pi N/(N+1) \to \pi$ in the formulas (\ref{fKfnm}), (\ref{GLnm}),
(\ref{vfd1}) and (\ref{vfd2}).  Let us recall the standard rules
of the transitions from the finite-difference operators to their
continuum counterparts:
\begin{eqnarray}
\sum \Delta \to \int dx; \label{transit1} \\
\frac{f(x_{n})-f(x_{n-1})}{\Delta} \to \frac{df}{dx}; \label{transit2} \\
\frac{f(x_{n+1})-2 f(x_{n})+f(x_{n-1})}{\Delta ^2} \to \frac{d^2 f
}{dx^2}.
 \label{transit3}
\end{eqnarray}

Now let us look at the Parseval's relation that takes, in the
continuum limit, its usual form for the infinite-dimensional
(Hilbert) space
\begin{eqnarray} \sum_{\mu =1}^{\infty}
\stackrel{\circ}{c}_{\mu }^{2}
\stackrel{\circ}{\varphi}(x,\stackrel{\circ}{E}_{\mu})
\stackrel{\circ}{\varphi}(y,\stackrel{\circ}{E}_{\mu})=
\delta(x-y),
\label{pars3}
\end{eqnarray}
and the same is for the new regular solutions $\varphi(x,E)$.

In the continuum case, we have $\varphi(0,E)=0, \enskip
\varphi'(0,E)=1$. Spectral weight factors are in that case, too,
the coefficients of proportionality between normalized
eigenfunctions and regular solutions. That is why they are also
referred to as norming constants since the multiplication by
$c_{\nu} = 1/\int_{0}^{\pi} \varphi^{2}(x,E_{\nu})dx$ turns
regular solution (at $E=E_{\nu}$) into the normalized one,
$$\Psi'(x,E_{\nu})|_{x=0} =c_{\nu}.$$

The expression for the transformed regular solutions has now the
following form (using (\ref{transit1}))
\begin{eqnarray} \varphi(x,E_{\nu})=\stackrel{\circ}{\varphi
}(x,E_{\nu})+ \int_{0}^{x} K(x,y)\stackrel{\circ}{\varphi }(y,E_{\nu})dy,
\label{phitr}
\end{eqnarray}
and similarly
\begin{eqnarray} \varphi(x,E)=\stackrel{\circ}{\varphi
}(x,E)+ \int_{0}^{x} K(x,y)\stackrel{\circ}{\varphi }(y,E)dy,
\label{phitr1}
\end{eqnarray}
 where $x \in [0,\pi]$. These formulas have just demonstrated that
the passage to the limit $\Delta \to 0$ does exist.  For the
kernel $K$ of the operator (\ref{phitr}) which transforms the
solutions  to the initial potential into the solutions to the new
one (generalized shift operator), we have the continuum analog of
Eq.  (\ref{GLnm}) -- the inverse problem equation proper:
\begin{eqnarray} K(x,y)+Q(x,y)+\int_{0}^{x} K(x,z)Q(z,y)dz=0,
\label{kq} \end{eqnarray} where the kernel Q is constructed from
the unperturbed functions with the old and new spectral parameters
[as in Eq.  (\ref{Q2})]:  \begin{eqnarray} Q(x,y)=\sum_{\nu
}c_{\nu}^{2}\stackrel{\circ}{\varphi }(x,E_{\nu})
\stackrel{\circ}{\varphi }(y,E_{\nu})- \sum_{\mu
}\stackrel{\circ}{c}^{2}_{\mu } \stackrel{\circ}{\varphi
}(x,\stackrel{\circ}{E}_{\mu}) \stackrel{\circ}{\varphi
}(y,\stackrel{\circ}{E}_{\mu}).  \label{Q3}
\end{eqnarray}

For the continuous coordinate, the expression (\ref{kp0pnm})
for $K$ has a similar form:  \begin{eqnarray} K(x,y) =
-\sum_{\nu }c_{\nu}^{2}\varphi (x,E_{\nu}) \stackrel{\circ}{\varphi
}(y,E_{\nu}) + \sum_{\mu }\stackrel{\circ}{c}^{2}_{\mu } \varphi
(x,\stackrel{\circ}{E}_{\mu}) \stackrel{\circ}{\varphi
}(y,\stackrel{\circ}{E}_{\mu}).  \label{kphi0phi} \end{eqnarray}

As we have just carried out the passage to the continuum limit in
solutions (\ref{fKfnm}), it is clear that such a limit exists for
the potential, too. In fact, the expressions for the potential
coefficients (\ref{vfd1}) and (\ref{vfd2}) were secondary with
respect to (\ref{fKfnm}), i.e. we always `extract' them from
Schr\"odinger equation using the information  about its solutions
(see the section above). As was shown, this procedure is
substantially based upon the completeness relation which stands
good for {\it any} $\Delta$, including the continuum case.
Moreover, we might use the continuum solution (\ref{phitr1}) and
the Parseval's relation (\ref{pars3}) for the continuum potential
to be derived. However, we choose the way of continuum passage in
(\ref{vfd1}) and (\ref{vfd2}). Another point is that the continuum
potential is {\it local}. Indeed, the u's and V in each row of the
discrete Sturm-Liouville operator are specified at the very
neighbour points $x_{n}$  and $x_{n+1}$ merging if we pass to the
continuum limit, which entails, in turn, superimposing the
potential coefficients at one point: $V+2 u$. The distinct feature
of the local potential is that the limiting equation must
determine it uniquely in contrast to (\ref{vfd1}) and (\ref{vfd2})
which, by themselves, could not uniquely specify u's and V's by
reason of a `non-local' character of the interaction in the
discrete case (we remember that there was required an additional
knowledge of $u(x_{N})$ at the boundary of the interval for the
uniqueness). But  for the Schr\"odinger equation with a local
potential,  it is well known that the potential always occur for
the unique  solution (with given boundary conditions) and vice
versa. Thus, we can beforehand anticipate an expression for a
unique specification of the local potential in the continuum case.

We shall now prove that, as $\Delta \to 0$, the equations
(\ref{vfd1}) and (\ref{vfd2}) go over, respectively, into
\begin{eqnarray}
\{V_{d}(x)-
\stackrel{\circ}{V}_{d}(y) + 2[u_{d}(x)- \stackrel{\circ}{u}_{d}(y)]\}
K(x,y) \nonumber \\ = \frac{\partial ^2}{\partial x^2}K(x,y) -
\frac{\partial ^2}{\partial y^2}K(x,y), \label{vdel20}
\end{eqnarray}
and \begin{eqnarray} \left\{ \begin{array}{l} {\tilde
V}_{d}(x) -  \stackrel{\circ}{V}_{d}(x) + {\tilde u}_{d}(x)
-\stackrel{\circ}{u}_{d}(x) = 2\frac{d}{dx}K(x,x) \\ \{{\tilde V}_{d}(x) -
\stackrel{\circ}{V}_{d}(x) + {\tilde u}_{d}(x)
-\stackrel{\circ}{u}_{d}(x)\} K(x,x) \\ =
\frac{\partial ^2}{\partial x^2}K(x,y)|_{y=x}
- \frac{\partial ^2}{\partial y^2}K(x,y)|_{y=x},
\end{array}
\label{vdel21}
\right.
\end{eqnarray}
where  $V_{d}(x) \equiv \lim_{\stackrel{\Delta \to 0}{m \to
\infty}}V(x_{m})$ and, analogously, $u_{d}(x) \equiv
\lim_{\stackrel{\Delta \to 0}{m \to \infty}}u(x_{m})$. The tilde
sign stands for the potentials obtained in passing to the limit of
continuous coordinate in the solutions of Eqs. (\ref{vfd2}).

 In
developing these equalities, it is useful to employ the diagonal
terms $K(x_{n},x_{n})$ such as $K(x_{n+1},x_{n})-K(x_{n},x_{n})
\sim O(\Delta)$. First of all, let us consider the term
$(K(x_{m+1},x_{m-1})-K(x_{m},x_{m-2}))/\Delta ^2$ in (\ref{vfd2}).
We add to and subtract from the expression in the numerator the
term $K(x_{m-1},x_{m-1})-K(x_{m},x_{m})$. Then
\begin{eqnarray*}
\frac{K(x_{m+1},x_{m-1})-K(x_{m},x_{m-2})}{\Delta ^2}
=\frac{K(x_{m+1},x_{m-1})+K(x_{m-1},x_{m-1})}{\Delta ^2} \\
-\frac{K(x_{m},x_{m}) +K(x_{m-1},x_{m-1})-K(x_{m},x_{m})
+K(x_{m},x_{m-2})}{\Delta ^2}=\zeta.
\end{eqnarray*}
Next, let us again add to and subtract from the new expression in
the numerator the term $2 K(x_{m},x_{m-1})$:
\begin{eqnarray*}
\zeta=\frac{K(x_{m+1},x_{m-1})-2 K(x_{m},x_{m-1})+K(x_{m-1},x_{m-1})}{\Delta ^2} \\
-\frac{K(x_{m},x_{m})-2 K(x_{m},x_{m-1})+K(x_{m},x_{m-2})}{\Delta ^2} \\
+\frac{K(x_{m},x_{m})-K(x_{m-1},x_{m-1})}{\Delta ^2}
\end{eqnarray*}
The first two lines in this expression are the second derivatives
with respect to the first and second argument of $K(x,y)$ (see
(\ref{transit3})). Hence, in the continuum limit they become
$$ \frac{\partial ^2}{\partial x^2}K(x,y)|_{y=x} -
\frac{\partial ^2}{\partial y^2}K(x,y)|_{y=x}.$$ The third
fraction diverges as $\Delta \rightarrow 0$: $\Delta ^{-1}
dK(x,x)/dx$. As a result we have
\begin{eqnarray}
\frac{K(x_{m+1},x_{m-1})-K(x_{m},x_{m-2})}{\Delta ^2}
\longrightarrow \frac{\partial ^2}{\partial x^2}K(x,y)|_{y=x} -
\frac{\partial ^2}{\partial y^2}K(x,y)|_{y=x} \nonumber \\ +
\Delta ^{-1} \frac{d}{dx}K(x,x). \label{lim1}
\end{eqnarray}
 Likewise, it is not difficult to obtain that
\begin{eqnarray}
\frac{K(x_{m+1},x_{m})-K(x_{m},x_{m-1})}{\Delta ^2} \nonumber \\
=\frac{K(x_{m+1},x_{m})-K(x_{m},x_{m})+K(x_{m},x_{m})-K(x_{m},x_{m-1})}{\Delta
^2} \nonumber \\
\longrightarrow  \Delta ^{-1} \frac{d}{dx}K(x,x),
\enskip \Delta \rightarrow 0. \label{lim2}
\end{eqnarray}

In equation (\ref{vfd1}) we see the finite-difference second
derivative in an explicit form. So in continuum case this equation
becomes (\ref{vdel20}). If we introduce $V(x) \equiv V_{d}(x)+2
u_{d}(x)$ the term in front of $K(x,y)$ is simply the difference
$V(x)-\stackrel{\circ}{V}(y)$. It is obvious that we introduced a
local limiting potential which results from the limiting merging
of V-diagonal and nearby u-diagonals.

Now let us multiply both sides of the equations of (\ref{vfd2})
for $n=m-1$ and $n=m$ by $\Delta$. We sum the resulting equations
and pass to the continuum limit. Then, by virtue of (\ref{lim1})
and (\ref{lim2}), we get to the first equation in (\ref{vdel21})
valid to within  $O(\Delta)$ (the multiplication by $\Delta$ has
removed the divergence associated with $\Delta ^{-1}$).

The last equation in (\ref{vdel21}) is not obvious. Indeed, one
would think that  the term $u(x_{m})$ must first be derived from
the recurrence procedure (\ref{vfd1}) and (\ref{vfd2}) and only
afterwards can the passage to the limit $\Delta \to 0$ be carried
out -- the procedure of a prodigious complexity. However, we find
a way out: we simply take the sum of non-diverging terms (taking
into account the expression (\ref{lim1})) in the right-hand side
of the equation (\ref{vfd2}), $n=m-1$ to be zero in the continuum
limit, i.e., we get  the last equation in (\ref{vdel21}). This by
no means contradicts the uniqueness of the sought limiting
potential. First, this provides the limiting (for $x=y$) equation
for $K(x,y)$ which must exist, obviously. Second, by continuity,
the factor $V_{d}(x)- \stackrel{\circ}{V}_{d}(y) + 2[u_{d}(x)-
\stackrel{\circ}{u}_{d}(y)]$ in front of $K(x,y)$ must coincide
with ${\tilde V}_{d}(x) - \stackrel{\circ}{V}_{d}(x) + {\tilde
u}_{d}(x) -\stackrel{\circ}{u}_{d}(x)$ when $x=y$. In other words,
that means that $V(x)=V_{d}(x)+2 u_{d}(x)={\tilde V}_{d}(x) +
{\tilde u}_{d}(x) + \stackrel{\circ}{u}_{d}(x)$, i.e. the
solutions of (\ref{vfd2}) go over, in the limit $\Delta \to 0$,
into {\it the same} local potential $V(x)$, which was beforehand
clear. Hence, with the new definition for $V(x)$, we have from
(\ref{vdel20}) and (\ref{vdel21}):
\begin{eqnarray} V(x) = \stackrel{\circ}{V}(x) + 2 \frac{d}{dx}K(x,x).
\label{vK} \end{eqnarray} This is the known result of recovering
potential in continuum case, which only now became reproducible
from a discrete mathematics.

The Eqs. (\ref{vdel20}) and
(\ref{vdel21}) can now be rewritten as
\begin{eqnarray} \left\{ \begin{array}{l}
\{V(x)- \stackrel{\circ}{V}(y)\} K(x,y)  \\ =
\frac{\partial ^2}{\partial x^2}K(x,y) -
\frac{\partial ^2}{\partial y^2}K(x,y)
\\
V(x)-\stackrel{\circ}{V}(x) = 2 \frac{d}{dx}K(x,x) \\
\end{array}
\label{vdel22}
\right.
\end{eqnarray}
This system (added by $K(0,0)=0$) represents the classical Goursat
problem [for determining $K(x,y)$] and its solvability follows
from well known theorems.

The orthogonalization can also be started from the last vector
$\varphi(\pi)$ with `number' $x=\pi$ at the right boundary of the
interval $[0,\pi]$.  Then, instead of solutions $\varphi(x)$, the
solutions $f(x)$ will be used such that $f(\pi)=0, \enskip
f'(\pi)=1$.  The corresponding inverse problem equations, that can
be associated with  the orthogonalization `from the right to the
left', have analogous form as Eqs. (\ref{phitr}), (\ref{kq}),
(\ref{Q3}) and(\ref{vK}), only with other integration limits and
different sign in front of the derivative in the expression for
$V(x)$:
\begin{eqnarray} f(x,E)=\stackrel{\circ}{f}(x,E)+
\int_{x}^{\pi} K(x,y)\stackrel{\circ}{f}(y,E)dy; \label{phitrr}
\end{eqnarray}
\begin{eqnarray} K(x,y)+Q(x,y)+\int_{x}^{\pi}
K(x,z)Q(z,y)dz=0; \label{kqr} \end{eqnarray}
\begin{eqnarray}
Q(x,y)=\sum_{\nu }\gamma_{\nu}^{2}\stackrel{\circ}{f}(x,E_{\nu})
\stackrel{\circ}{f}(y,E_{\nu })-
\sum_{\mu }\stackrel{\circ}{\gamma}^{2}_{\mu }
\stackrel{\circ}{f}(x,\stackrel{\circ}{E}_{\mu})
\stackrel{\circ}{f}(y,\stackrel{\circ}{E}_{\mu});
\label{Q3r}
\end{eqnarray}
\begin{eqnarray}
V(x)= \stackrel{\circ}{V}(x) - 2 \frac{d}{dx}K(x,x).
\label{vKr}
\end{eqnarray}
Here, the symbol $\gamma_{\nu}$ stands for the spectral weight
factor which is analog of $c_{\nu }$. The only discrepancy is
that the  $\gamma_{\nu}$ characterizes the behaviour of eigenfunction
at the right boundary:  $$\Psi(x,E_{\nu})=
\gamma_{\nu} f(x,E_{\nu}), \enskip \gamma_{\nu
}=\frac{d}{dx}\Psi(x,E_{\nu})|_{x=\pi}.$$

At last, let us mention about the eigenvalue inverse problem for
Schr\"odinger equation added by boundary conditions of arbitrary kind:
\begin{eqnarray} \Psi'(0) - g \Psi(0) = 0, \enskip
\enskip \enskip \Psi'(\pi) + G \Psi(\pi) = 0.  \end{eqnarray}
Here we also have analogous inversion equations, and as
spectral weight factors  there appear the values of
corresponding eigenfunctions at the interval edges:
$c_{\nu }=\Psi(0,E_{\nu})$ or $\gamma_{\nu}=\Psi(\pi,E_{\nu})$.

\section{Conclusions}

In the present paper we carried out the derivation of main
formulas of the inverse eigenvalue problem on the base of its
discrete approximation. Several statements of that problem are
developed by now, we selected such a statement in which it is
possible to reproduce in a maximally straightforward way the
future structure of the limiting inversion procedure:  the
transition from a known system to the system with given spectral
data (eigenvalues plus norming constants) but with unknown
potential to be recovered. The off-diagonal elements are
introduced into the matrix Sturm-Liouville operator
(three-diagonal matrix), which is consistent (in contrast to
previous works) with the problem statement involving this double
set of spectral parameters. At last, in comparison with usual
derivation of the continuum inversion equations, our development
seems to be none the more complicated. At the same time, the
reader acquires the ability to track in more detail additional
aspects of the formalism, in particular to look upon the operator
transformation realizing the recovering procedure as the
orthonormalization of the operator eigenvectors.

\end{document}